\documentclass[runningheads]{llncs}
\usepackage[T1]{fontenc}
\usepackage{url}
\usepackage{multirow}
\usepackage{tabularx}
\usepackage{booktabs}
\usepackage{makecell}

\usepackage{graphicx}
\begin{document}
\title{An Empirical Analysis of ReDoS Vulnerabilities and ReDoS Detection Tools}
\author{N'Zolieh Ismaël Mahassadi\inst{1} \and
Raphaël Khoury\inst{1} \and
Justin Vallé\inst{1} \and 
Abdelwahab Hamou-Lhadj\inst{2}} 
\authorrunning{N. Mahassadi et al.}
\institute{Université du  Québec en Outaouais, Gatineau, Canada
\email{\{mahy04,raphael.khoury,valj27\}@uqo.ca}\and 
Concordia Universiy, Montréal, Canada
\email{wahab.hamou-lhadj@concordia.ca }}

\maketitle              % typeset the header of the contribution
\begin{abstract}
ReDoS vulnerabilities are a type of denial of service software weakness that occurs when a regex is used to validate user-supplied input.  In some cases, the regex matching process can take exponential time, leading to a denial of service. In this study, we examine and compare the effectiveness of five publicly-available regex detection tools, and one regex correction tool, using three datasets. We further perform an empirical analysis of all ReDoS vulnerabilities reported to the NVD database in order to understand how they differ from non-ReDoS vulnerabilities and glean insights about this type of weakness. We find that ReDoS vulnerabilities are becoming more prevalent and are much more likely to be exploited than non-ReDoS vulnerabilities. We further find that detection tools exhibit substantial disagreement on whether or not a given regex is vulnerable. 

\end{abstract}

\section{Introduction}
\label{sec:intro}
ReDoS vulnerabilities (for regex denial of service) are a category of vulnerabilities that may be present when a regex is used to validate the format of user-supplied input. In some cases, the regex validation process can take a supralinear amount of time based on the length of the input, leading to a denial-of-service.

Two conditions must be met for a ReDoS attack to be possible. First, the attack is only possible if the regex validation engine uses back-tracking\cite{first}. Some regex validation libraries are immune to the attack, regardless of the regex pattern being matched. Secondly, the attack occurs if the regex exhibits one of several particular patterns, termed \textit{evil patterns}.    

If both conditions are met, and the regex is used to validate an input submitted by a malicious adversary, it can submit a specially crafted input whose validation by the regex takes exponential (or a large polynomial) time. 

The literature on this topic is not in agreement regarding the exact definition of these patterns. The definition proposed by Li et al.\cite{Redoshunter} seems to be the most thorough. They identify 5 distinct evil patterns, three of which incur the possibility of a supralinear evaluation time, and two others that risk large polynomial blow-ups. This classification is not universally accepted; consequently, some regex detection tools can recognize only a subset of vulnerable patterns. 

Perhaps because of the uncertainty about which patterns are problematic, this category of attack is becoming more prevalent. Furthermore, for the same reason, the various tools that developers may rely on to determine if the regex they wish to use exhibits an evil pattern are often not in agreement.    

In this study, we perform two experiments that shed light on this category of vulnerability, providing actionable information for developers and security professionals. 

The contributions of this paper are as follows:
\begin{enumerate}
    \item We perform an empirical analysis of over 400 ReDoS CVEs, highlighting how they differ from CVEs that report on other classes of vulnerabilities. 

    \item We evaluate the effectiveness of five ReDoS vulnerability detection tools and one ReDoS correction tool, using three datasets of regexes. Finding that ReDoS detection tools are often not in agreement, we propose a multi-factor metric for determining if a regex is vulnerable, and argue that our experimental results support the use of this metric.

    \item We make two datasets available to the software security community: first, a dataset of 490 ReDoS CVEs, which is useful for research purposes since ReDoS CVEs are often not labeled as such. Second, a labeled version of RegexLib dataset, indicating whether each regex is vulnerable or not according to the above-mentioned  multi-factor metric. 
\end{enumerate}

The remainder of this paper is organized as follows: Section \ref{sec:related} surveys related works and provides additional background information about ReDoS vulnerabilities. Section \ref{sec:desc} describes the datasets relied upon for this paper and provides an outline of our methodology. Section \ref{sec:emp} shows the results of our analysis of ReDoS CVEs. Section \ref{sec:prac} provides the results of our experiments with ReDoS detection tools. Concluding remarks are given in Section \ref{sec:conclu}.

 \section{Related Works} \label{sec:related}

ReDoS vulnerabilities have been the topic of fewer studies than other, more common classes of vulnerabilities. Nonetheless, two large scale empirical studies found them to be surprisingly prevalent. Davis et al. \cite{Davis2018impact} found that 1\% of over 400,000 regexes extracted from a similar number of python modules are vulnerable, while  Staicu et al. \cite{Staicu2018Freezing} found that 339 of the most popular 2,846 websites were vulnerable.  Both authors use a definition of vulnerable regex that may overlook some of the malicious patterns defined by Li et al. \cite{Redoshunter}.

A number of studies develop tools to automatically detect whether a given regex is vulnerable or not \cite{Saferegex,Rescue,Redoshunter,revealer,RAT}. These span static, dynamic, and hybrid methods.  We review these tools in the next section and evaluate their effectiveness in Section \ref{sec:prac}.   Another line of research seeks to re-write or correct regexes to avoid supralinear patterns \cite{RFixer,scalpel}. Amongst these methods, we focus on RegexScalpel \cite{scalpel}, which proceeds through a series of modifications that may alter the regex's semantics. 
 
\section{Description of the Datasets and Empirical Setup}
 \label{sec:desc}

\subsection{Datasets}
%https://github.com/JustinVallee/CVE-ReDoS-Analysis/blob/main/data_redos/data_cvss/redos_cvss.csv
\subsubsection{ReDoS CVE Dataset}
We searched the NVD database for CVEs that refer to ReDoS vulnerabilities. CWE-1333 : ``Inefficient Regular Expression Complexity", refers specifically to this class of vulnerability. As of September 2025, the NVD contains 341 entries marked with this CWE.  However, a manual inspection found that many CVEs describe ReDoS vulnerabilities but either have a different CWE label or no label. To be thorough, we aggregated to our dataset any vulnerability whose description contains any one of the following terms: ``ReDoS, regular expression denial of service, Regex Denial of Service, catastrophic backtracking, inefficient regex, inefficient regular expression, regex performance, regex dos." %This process added 141 CVEs, for a total of 490 entries. We will return to the question of the CWE labels of CVEs for ReDoS vulnerabilities in Section \ref{sec:emp}.

In section \ref{sec:emp}, we dive into this dataset to understand how regex vulnerabilities differ from non-regex vulnerabilities.  

\subsubsection{RegexLib}
RegexLib\footnote{\url{https://regexlib.com/}}  is a freely available online database of over 3500 regexes that validate a wide variety of useful patterns.  Research \cite{reuse} shows that when developers need a regex for a commonplace purpose, they are more likely to 'borrow' a regex from a repository such as RegexLib than to create one themselves. Research further shows that programmers often insert regexes into their code without  completely understanding them. As a consequence, it is likely that developers unwittingly include regexes that are vulnerable to ReDoS attacks in their code. For this project, we obtained the RegexLib data from RXXR2\footnote{\url{https://github.com/superhuman/rxxr2/tree/master.}}, a sanitized and formatted set created from RegexLib. From an initial dataset of 3514 regexes, we excluded  661 regexes because they were either malformed, ambiguous, or trivially simple, and preserved 2853 regexes for analysis.

\subsubsection{CVEFixes}
CVEFixes \cite{cvefixes} is a dataset of 5,000 vulnerable and patched code extracted from the NVD database. From this dataset, we selected those that referred to ReDoS vulnerabilities and extracted all regexes from both the vulnerable and patched versions of the program.  In what follows, we respectively refer to regexes extracted from vulnerable code and patched code, the Regex\_before and Regex\_after datasets. We apply the same suite of regex testing tools to both datasets, gleaning insights about the prevalence of this vulnerability and the effectiveness of countermesures. 

After excluding malformed regexes, the Regex\_before dataset contains 379 entries, while the  Regex\_after  dataset contains 391 entries. It is important to stress that the regexes in the former dataset are not necessarily the vulnerable regexes that incurred the underlying vulnerability, and likewise, the regexes in the latter set are not necessarily patched versions of these regexes (though that is surely the  case for some regex pairs). Rather, all regexes present in the initial (resp. patched) version of the code are present  in  Regex\_before (resp. Regex\_after), and the two sets contain some duplicates. The goal was only to generate a fair number of real-life regexes to test ReDoS detection tools. 

In section \ref{sec:prac}, we employ a suite of ReDoS detection tools to determine if the regexes present in the RXXR2  Regex\_before and Regex\_after datasets  contain these vulnerable regexes.  The dataset of CVEs is present on one of the author's Github\footnote{\url{https://github.com/JustinVallee/CVE-ReDoS-Analysis/blob/main/data_redos/data_cvss/redos_cvss.csv}}.

%\footnote{\url{https://github.com/JustinVallee/CVE-ReDoS-Analysis/blob/main/data_redos/data_cvss/redos_cvss.csv}} 

\subsection{Analysis tools}
ReDoS detection tools are broadly categorized as static tools, which examine a regex to discern the presence of an evil pattern, and dynamic tools, which attempt to uncover the presence of a vulnerability by randomly generating  and testing possible inputs. Static analyzers are prone to false positives, while dynamic ones may miss vulnerable regexes. 
We create an automated pipeline that applies five different regex detection tools to every regex listed in a plain text file. These tools are:

\begin{itemize}
  \item \texttt{Saferegex} \cite{Saferegex} : A static analysis tool developed by Microsoft that detects malicious regexes using a process similar to model checking;
 % \item ReDoS Abstract Tester (RAT)\cite{rat} : A static analysis tool; 
 \item \texttt{Rescue} \cite{Rescue}	: A dynamic testing tool that generates instances of malicious input using a gray-box approach. Rescue may either mark a regex as vulnerable or invulnerable, or timeout (T.O.), an indication that the regex is likely vulnerable; 
\item \texttt{ReDoSHunter} \cite{Redoshunter}: A hybrid (static and dynamic) tool that detects up to 5 types of vulnerable patterns (EOD, EOA, POA, NQ and SLQ), defined by the author; 
\item \texttt{Revealer} \cite{revealer}: A hybrid (static and dynamic) tool that classifies vulnerable regexes as either susceptible to polynomial (poly.) or exponential (expo.) exploitation. 
\item \texttt{RAT} \cite{RAT} : A static analyzer developed by Parolini et al. RAT is a fast analyzer that exhibits a very low rate of false positives ( as low as 0.07\% in a study by the author). 
\end{itemize}

After applying these tools, we employed RegexScalpel \cite{scalpel} to generate a corrected version of each vulnerable regex. RegexScalpel is a tool developed by Li et al. to correct potentially vulnerable regexes. Only one other regex correction tool is freely available,  \cite{RFixer}. However, we excluded this tool because it cannot handle regexes that contain non-classical features such as backreferences.

It is important to stress that RegexScalpel does not always produce a corrected regex that is semantically equivalent to the original regex. In fact, in many cases, the tool 'fixes' a regex by imposing a hard-coded limit on the number of iterations of a quantifier (using the \{value\} construction).  

Finally, in order to evaluate the effectiveness of the suggested correctives, we analyze every corrected regex with the same suite of detection tools. In cases when the regex was still found to be vulnerable, this was followed by a second correction step.

%%%%%%%%%%%%%%%%%%%%%%%%%%%%%%%%%%%%%%%%%%%%%%%%%%%%%%%%%%%%%%%%%%%%%%%%%%%%%%%%%%%%%%%%%%%%
\section{Empirical Analysis}
 \label{sec:emp}
 In this section, we perform an empirical analysis of 490 CVEs that report on ReDoS vulnerabilities (referred to below as ReDoS CVEs) and compare them to the remainder (215,085) of CVE entries  (non-ReDoS CVEs) to understand how ReDoS vulnerabilities are distinct. 

\subsection{Prevalence}
\begin{figure}[]
\centering
\includegraphics[width=\columnwidth]{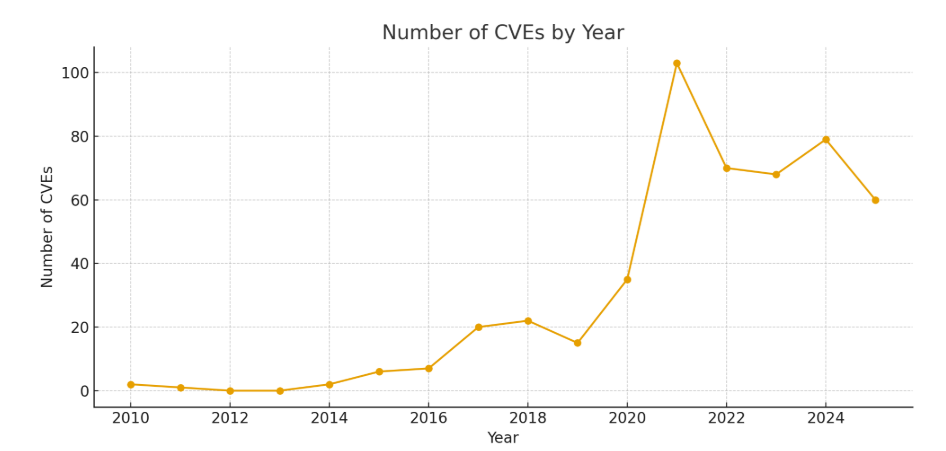}
\caption{Number of ReDoS CVEs by year.}
\label{fig:years}
\end{figure}

Figure \ref{fig:years} shows the number of ReDoS vulnerabilities reported to the NVD each year since 2010.  

Despite the fact that the NVD database has been maintained since 2002, the earliest ReDoS vulnerability only dates back to 2010. Furthermore, this category of attack was comparatively rare, numbering fewer than 20 annual occurrences until 2020. That year, the number of reported vulnerabilities suddenly spiked to 103, a surge that has been observed in relation to other types of vulnerabilities \cite{khoury2025mining}, and seems to coincide with an effervescence of malicious activity during the confinement caused by the Covid pandemic. Nonetheless, in the years that followed, ReDoS vulnerabilities have consistently been reported to the NVD at a rate of approximately 60 per year, indicating a continued interest on the part of malicious adversaries.

\subsection{CVSS scores}
Each CVE entry is associated with a CVSS score, which measures the severity of the vulnerability according to multiple factors. The CVSS score exists in multiple versions: namely, versions 2.0, 3.1, 3.0, and 4.0. In some cases, multiple CVSS scores are provided. 

For the analysis in this subsection, we set aside all CVEs that contain only a CVSS score of version 2.0 or 4.0. In case both CVSS versions 3.1 and 3.0 are present, we prioritize the former over the latter. This yielded a subset of  464 ReDoS vulnerabilities and  215,085 non-ReDoS vulnerabilities. Only 26 ReDoS CVEs did not have either a CVSS version 3.1 or 3.0, and comparison of CVSS scores between different versions would have been impractical. 

Results are given in Table \ref{tab:cvss}. As can be seen, ReDoS vulnerabilities differ from non-ReDoS vulnerabilities for every CVSS attribute.   Unsurprisingly, ReDoS attacks are overwhelmingly Network attacks (97.64\%), require no privileges (84.48\%), and needs no user interaction (96.12\%). ReDoS vulnerabilities have low attack complexity (96.77\%), but also low impact. Overall, they have a somewhat lower average base score (6.88 v. 7.11 for non-ReDoS vulnerabilities). 

Previous research \cite{iotMalware} showed that vulnerabilities exhibiting particular combinations of High impact and low attack complexity  (Called HiLac vulnerabilities) are more likely to be exploited.  We examined our dataset to determine whether the prevalence of HiLac vulnerabilities differed between ReDoS and non-ReDoS vulnerabilities. For this analysis, we considered a vulnerability to have a high impact if it's impact score is higher than 7.0 (note that this includes critical vulnerabilities with an impact score between 9.0 and 10). 

We found HiLac ReDoS to be rare: only 0.43\% of ReDoS vulnerabilities are HiLac, versus 42.33\% for non-ReDoS vulnerabilities, a result that may lead security analysts to de-prioritize ReDoS vulnerabilities---mistakenly, as we will see in the next section.

% \usepackage{multirow}
\begin{table*}[]\caption{Comparaision of CVSS Attribute values between ReDoS and non-ReDoS Vulnerabilties}
\centering
\begin{tabular}{|l|l|l|l|l|}
\hline
CVSS Metric                         & Value            & ReDoS                & Non-ReDoS     & Chi-Sq Value                                                                     \\ \hline
\multirow{2}{*}{User Interaction}   & None             & 140,978 (65.55 \%)   & 446(96.12\%)  & \multirow{2}{*}{\begin{tabular}[c]{@{}l@{}}2.49e-43\\    \\ (diff)\end{tabular}} \\ \cline{2-4}
                                    & Requiered        & 74,107  (34,45\%)    & 18(3.88\%)    &                                                                                  \\ \hline
\multirow{4}{*}{Attack Vector}      & Adjacent Network & 5,222  (2.43\%)      & 2 (0.43\%)    & \multirow{4}{*}{4.686e-30 (diff)}                                                \\ \cline{2-4}
                                    & Local            & 49,927   (23.21\%)   & 9(1.94\%)     &                                                                                  \\ \cline{2-4}
                                    & Network          & 15,7859 (73.39\%)    & 453(97.63\%)  &                                                                                  \\ \cline{2-4}
                                    & Physical         & 2,077   (0.97\%)     & 0(0.0\%)      &                                                                                  \\ \hline
\multirow{2}{*}{Attack Complexity}  & Low              & 13,770  (6.4\%)      & 15(3.23\%)    & \multirow{2}{*}{0.007 (diff)}                                                    \\ \cline{2-4}
                                    & High             & 201,315 (93.6\%)     & 449(96.77\%)  &                                                                                  \\ \hline
\multirow{3}{*}{Privilege Required} & High             & 18,211     (8.47\%)  & 5(1.08\%)     & \multirow{3}{*}{1.17e-28   (diff)}                                               \\ \cline{2-4}
                                    & Low              & 70,260     (32.67\%) & 67(14.44\%)   &                                                                                  \\ \cline{2-4}
                                    & None             & 126,614    (58.87\%) & 392(84.48\%)  &                                                                                  \\ \hline
\multirow{5}{*}{Base Score}         & 0-2              & 51 (0.02\%)          & 0 (0.0\%)     & \multirow{5}{*}{4.87e-03 (diff)}                                                 \\ \cline{2-4}
                                    & 2-4              & 4,637     (2.16\%)   & 8 (1.72\%)    &                                                                                  \\ \cline{2-4}
                                    & 4-6              & 55,828  (25.96\%)    & 87 (18.75\%)  &                                                                                  \\ \cline{2-4}
                                    & 6-8              & 94,717 (44.04\%)     & 367 (79.09\%) &                                                                                  \\ \cline{2-4}
                                    & 8-10             & 59,847   (27.83\%)   & 2 (0.43\%)    &                                                                                  \\ \hline
\end{tabular}\label{tab:cvss}
\end{table*}

\subsection{Exploitation and Exploitability}
For the purpose of this study, we distinguish exploitation, defined as the determination that a vulnerability has been exploited in the wild, from exploitability \cite{birthmark}, a metric that measures the likelihood that a vulnerability will eventually be exploited. 
Unfortunately, reliable data on exploitation is scarce. The most reliable data source is the Known Exploited Vulnerabilities (KEV) Catalog , maintained by the Cybersecurity and Infrastructure Security Agency. None of the ReDoS vulnerabilities we compiled are included in this listing. However, in previous research \cite{khoury2025mining}, we compiled a list of 5,000 vulnerabilities for which there was evidence of exploitation from any one of four sources  : the  KEV catalog and data shared by three industrial partners: Secureworks, Greynoise, and  ClamAV. This list captures vulnerabilities published between 2019 and 2023. The ReDoS dataset contains 286 vulnerabilities published during this time period, of which 277 (97\%) are listed. If these results bear out, it means ReDoS vulnerabilities may be the category of attacks most likely to be exploited by malicious adversaries  \footnote{This dataset is available here: https://zenodo.org/records/14026455}.

Previous research has identified a number of factors that correlate with exploitation, allowing for the computation of metrics that reflect the likelihood of exploitation, such as EPSS. Amongst the data that are most highly correlated with exploitation, we note the number of references present in the CVE report, the presence of references tagged ‘exploit’, the presence of references to the .gov top-level domain, as well as references to Metasploit and exploit-bd \cite{jacobs2021exploit}. A small number of highly targeted vendors are also correlated with a higher exploitation rate. We will return to the analysis of vendors in the next subsection.

Exploitability metrics generally provide inconclusive signals regarding the  likelihood of exploitation of ReDoS vulnerabilities.  On the one hand, ReDoS CVEs did exhibit slightly  more references  (avg 3.54, median 3.0) than non-ReDoS CVEs (avg 2.52, median 2.0). A Mann Whitney test shows this difference to be statistically significant (p-value = 1.741e-4), though very small.  

ReDoS CVEs are twice as likely as non-ReDoS CVEs (47.4\%  v. 26.2\%, a factor of 1.81) to have associated exploit code or POC, measured by the presence of a reference with the tag 'Exploit'. This is a significant indicator of the potential for exploitation. 

On the other hand, None of the  ReDoS CVEs  have references to URL in the .gov domain.  Likewise, none of the ReDoS CVEs have references to URL in the .gov domain, vs 4790 (2.2\%) for non-ReDoS  CVEs. The same holds for references to exploit-db  v. 4134 (1.9\%for non-ReDoS CVEs  and to metasploit, 131(0.06\%) for non-ReDoS CVEs.

Another important measure of exploitability is the EPSS score \cite{jacobs2021exploit}, which is computed daily and measures the likelihood that a CVE will be exploited in the subsequent 30 days. Since the EPSS score is computed daily, we looked at the peak score, i.e., the highest value ever assigned to a CVE as an indicator of how likely it is that this vulnerability would ever be exploited. ReDoS vulnerabilities average 0.04629 (std. 0.0859) versus 0.07261 (std. 0.16131) for non-ReDoS vulnerabilities, a difference that was found to be not statistically significant using a Mann-Whitney U test ( p = 0.2671). 

%Exploited CVEs Average EPSS Peak Value: 0.09068 
We also looked at how many days elapse between the initial publication of a vulnerability and the moment it reaches it's peak EPSS score. The peak occurs 498.40 days after publication (std. 586.34) for ReDoS vulnerabilities, vs. 698.30 days (std. 766.08) for non-ReDoS vulnerabilities. This difference is statistically significant (p = 1.11e-07) and indicates that malicious adversaries are quicker to develop attack code for ReDoS vulnerabilities, an indication of the ease with which such vulnerabilities can be exploited. 

The dissonance between the inconclusive indicators of exploitability and the very high rate of exploitation observed in our dataset is a powerful indication of the limitations of these metrics.    

\subsection{CPE}
\begin{figure}[]
\centering
\includegraphics[width=\columnwidth]{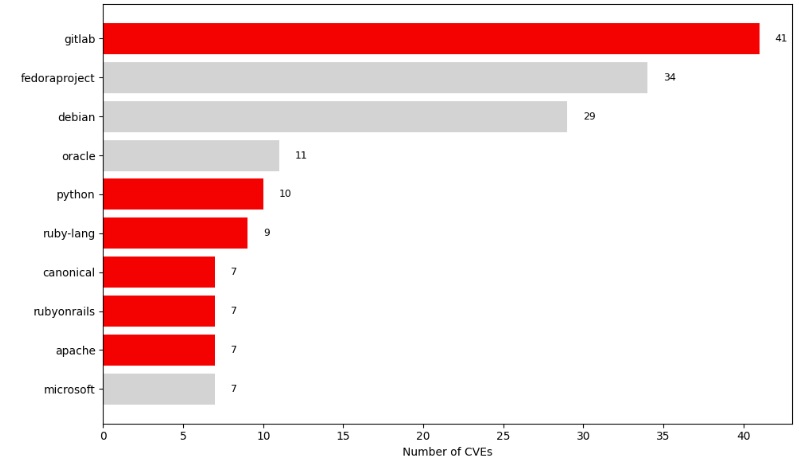}
\caption{Vendors for whom ReDoS vulnerabilities are most commonly reported.}
\label{fig:cpe}
\end{figure}

The CPE (Common Platform Enumeration) is a section of the CVE report that  lists the vendors and products affected by the corresponding vulnerability.  Figure \ref{fig:cpe} reports the top 10  most common Vendors for ReDoS  CVEs. Vendors that are also present in the top ten most common vendors for non-ReDoS vulnerabilities are highlighted in red, while vendors present only in the ReDoS top 10 are highlighted in gray. Note that a single CVE may be related to multiple vendors. 

Two observations are discernible from an analysis of this figure. First, Ruby programs are preponderant in ReDoS CVEs. This may be due to the use of this language in web frameworks. Note that GitLab is coded in Ruby.
Second, we observe that 4 of the top ten vendors, namely, Fedora Project, Debian, Oracle, and Microsoft—also appear on a list of vendors whose vulnerabilities are most commonly chosen for exploitation by malicious adversaries, as compiled in a previous study \cite{birthmark} . Python is the vendor that is most overrepresented in ReDoS CVEs, compared with non-ReDoS vulnerabilities. Interestingly, since March 2023, Ruby v. 3.2 incorporates an improved regex matching algorithm that makes ReDoS attacks impossible \cite{ruby}.

\subsection{CWE}
\begin{figure}[]
\centering
\includegraphics[width=\columnwidth]{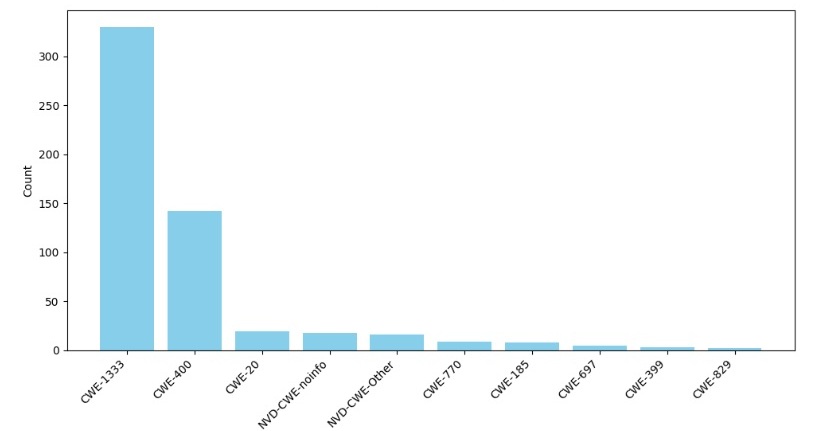}
\caption{Most common CWE for ReDoS vulnerabilities.}
\label{fig:cwe}
\end{figure}

As mentioned earlier, ReDoS vulnerabilities can be assigned one of several possible CWE. The most common CWE is CWE-1333 (330 CVEs), which specifically describes the ReDoS vulnerability. Other CWEs used to describe this vulnerability include: CWE-400 : 'Uncontrolled Resource Consumption' (142), CWE-20: 'Improper Input Validation' (19), CWE-770 :'Allocation of Resources Without Limits or Throttling' (9), CWE-185 : 'Incorrect Regular Expression' (8) , CWE-697: 'Incorrect Comparison' (5), CWE-399: 'Resource Management Errors' (3) and CWE-829 : 'Inclusion of Functionality from Untrusted Control Sphere '(2).   Figure \ref{fig:cwe} reports these results.  We note in passing that the use of several of these CWE labels contravenes with Mitre’s best practice guidelines \footnote{\url{https://cwe.mitre.org/documents/cwe_usage/guidance.html}} since they are not Base type CWEs. 

18 CVEs are assigned the label NVD-CWE-noinfo, while 16 are assigned the label NVD-CWE-Other.
This result confirms the ambiguity present in assigning CWEs to vulnerabilities, as noted by other researchers \cite{wu2015they}. 

For this analysis, in cases where more than one CWE was assigned to a given CVE, we prioritized the CWE assigned by NVD, followed by a CWE assigned by DHS, over CWEs assigned by industry partners.

%***************************************************************************
\section{Experimental Analysis}\label{sec:prac}
\subsection{Methodology}
In this section, we use a suite of five ReDoS detection tools to determine how common ReDoS vulnerabilities are in each of the three datasets of regexes constructed for this study and how effective publicly available tools are. Of the five tools tested, only ReDoSHunter includes a formal definition of the patterns deemed to be vulnerable, making a theoretical comparison of their capabilities difficult. Instead, we evaluate each regex in the datasets using each of the five tools through an automated pipeline.  All the code is available on our Github repository\footnote{\url{https://github.com/ISMAELYVANN/ReDoS-Evaluation-Pipeline/tree/main}} . Results are provided in table \ref{tab:results}.

%\footnote{\url{https://github.com/ISMAELYVANN/ReDoS-Evaluation-Pipeline/tree/main}}

As mentioned above, we analyze regexes from two sources, the RegexLib (via the RXXXR2 project)  and CVEfixes.  The regexes are first sanitized to eliminate malformed regexes, uniformize encoding, and ensure compatibility with the evaluation tools. 

The sanitized regexes are then sent to the five regex detection algorithms listed in section \ref{sec:desc}. Evil regexes are corrected using regexsclapel, and the corrected regexes are evaluated a second time using the same 5 tools in order to assess the effectiveness of the correction. 

\begin{figure}[]
\centering
\includegraphics[scale=.4]{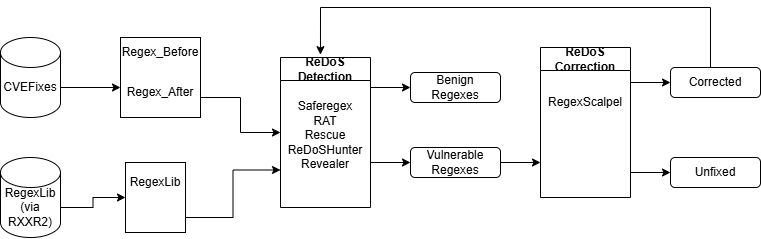}
\caption{Processing Pipeline}
\label{fig:pipeline}
\end{figure}

RegexScalpel can generate multiple possible corrections for a single vulnerable regex, in which case a single candidate is selected at random for the purposes of the analysis.  This is illustrated in figure \ref{fig:pipeline}.  

\subsection{Results and Analysis}

The results of the evaluation are presented in Table \ref{tab:results}. 
\begin{table*}[] \caption{Vulnerable Regexes according to each Detection tool}
\begin{tabular}{|l|l|l|l|ll|lllll|ll|l|}
\hline
\multirow{2}{*}{Dataset} & Total   & \multirow{2}{*}{Safe-} & \multirow{2}{*}{RAT} & \multicolumn{2}{l|}{Rescue}       & \multicolumn{5}{l|}{ReDoSHunter}                                                                               & \multicolumn{2}{l|}{Revealer}      & \multirow{2}{*}{Any} \\ \cline{5-13}
                         &  &               regex             &                      & \multicolumn{1}{l|}{Vuln.} & T.O. & \multicolumn{1}{l|}{EOD} & \multicolumn{1}{l|}{EOA} & \multicolumn{1}{l|}{POA} & \multicolumn{1}{l|}{NQ} & SQL & \multicolumn{1}{l|}{Poly} & Expo &                      \\ \hline
 R. before            & 323     & 62                         & 3                    & \multicolumn{1}{l|}{27}    & 120  & \multicolumn{1}{l|}{6}   & \multicolumn{1}{l|}{6}   & \multicolumn{1}{l|}{96}  & \multicolumn{1}{l|}{2}  & 29  & \multicolumn{1}{l|}{17}    & 7     & 61                   \\ \hline

R. after             & 330     & 41                         & 0                    & \multicolumn{1}{l|}{10}    & 125  & \multicolumn{1}{l|}{2}   & \multicolumn{1}{l|}{0}   & \multicolumn{1}{l|}{75}  & \multicolumn{1}{l|}{0}  & 28  & \multicolumn{1}{l|}{7}     & 0     & 34                   \\ \hline
RegexLib                     & 2853    & 608                        & 74                   & \multicolumn{1}{l|}{211}   & 680  & \multicolumn{1}{l|}{8}   & \multicolumn{1}{l|}{48}  & \multicolumn{1}{l|}{284} & \multicolumn{1}{l|}{26} & 151 & \multicolumn{1}{l|}{80}    & 118   & 440                  \\ \hline
\end{tabular} \label{tab:results}
\end{table*}

\begin{figure}[]
\centering
\includegraphics[scale =.5]{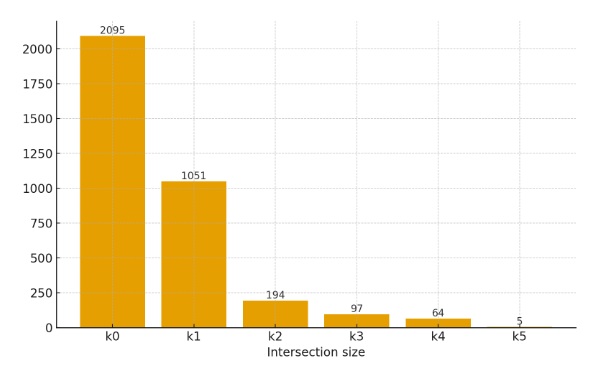}
\caption{Number of Regexes flagged as vulnerable by 0, 1, 2, 3, 4 and all 5 tools.}
\label{fig:bar}
\end{figure}

The tools exhibit greater disagreement than agreement, underscoring their complementarity and the necessity for a multi-tool study. Indeed, Figure\ref{fig:bar}  shows the number of regexes for which a single tool uncovered a vulnerability, as well as how often two (or more) tools achieved a concordant verdict that the regex is vulnerable.
As can be seen, 74\% of vulnerable regexes are flagged by a single tool (1051/1411), and in only 5 vulnerable cases do we have agreement for all five tools. 

The most common vulnerable pattern, according to the ReDoSHunter tool, consists of two or more juxtaposed quantifiers that share common valid sequences (e.g. [a-z]*[\textbackslash W]*).  
Rescue detects a very high number of timeouts, which are indicative of a potential vulnerability but may also reflect a false positive.  In fact, the tool returns more timeouts than definitive vulnerable verdicts for all three datasets. RAT remains very conservative, only flagging a small number of regexes as vulnerable. Nonetheless, it still manages to catch a number of vulnerable regexes that no other tool deems vulnerable. This is a concerning result, given that this tool is specifically designed to avoid false positives. 

Also concerning is the fact that 15\% of regexes present in the RegexLib dataset are marked as vulnerable by at least one of the tools. As discussed above, developers often reuse regexes taken verbatim from libraries, rather than creating them anew. Furthermore, even regexes that accomplish simple tasks, such as validating the format of a date, can be too complex to be easily understood by novice programmers. Consequently, the practice of 'borrowing' regexes from public libraries should be seen as imprudent.  The lack of awareness of ReDoS attacks only adds to the risks. The reduce this risk in the future, a labelled version of the RegexLib dataset, indicating which regexes are vulnerable, is provided on the author's Github repository.

\begin{figure}[]
\centering
\includegraphics[width=\columnwidth]{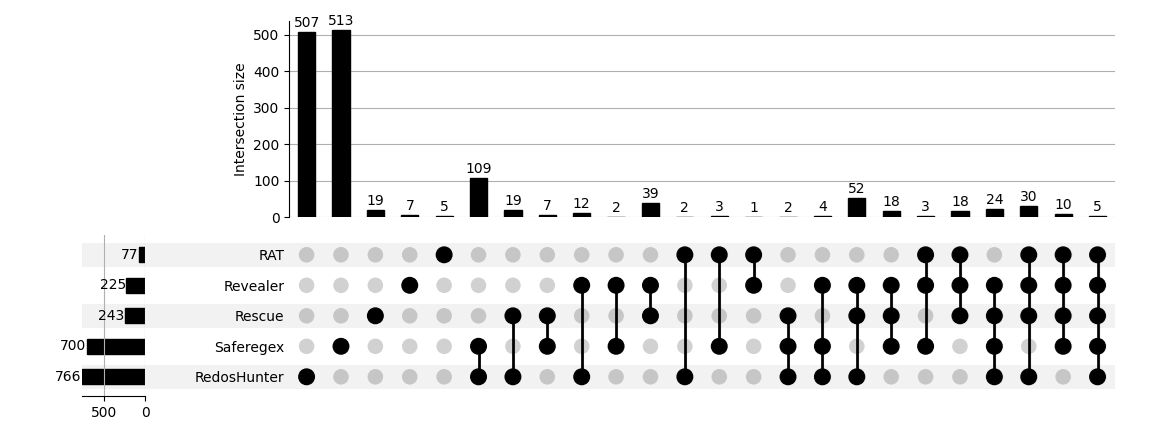}
\caption{Upset Plot of Regexes detected as vulnerable by each tool.}
\label{fig:upset}
\end{figure}

Figure \ref{fig:upset} shows an upset plot that highlights the intersection between each of the five tools under consideration. This upset chart is an alternative to the Venn Diagram that is more suited to the analysis of multiple sets. 

In the upset chart, the vertical bar represents the percentage of regexes marked as vulnerable by each tool (uniquely) as well as by each intersection of two, three, four, or five tools. The first five bars represent the percentage of regexes marked as vulnerable solely by one of the five tools, with all other tools returning a  negative result (i.e., an intersection of one). The black dots under each bar indicate which tools participate in the corresponding intersection. The horizontal bars on the left represent the percentage of all regexes marked as vulnerable by each tool.  

From this visualization, we can glean further insights: The bulk of  regexes deemed vulnerable are flagged by one of two tools:  Saferegex and RedosHunter. Furthermore, these two tools exhibit a low degree of agreement with each other and with other tools.  This result indicates that either  existing tools are inadequate, or they exhibit a high rate of false positives.   In the case of ReDoSHunter, such an outcome is expected since this tool is capable of detecting vulnerable patterns not covered by any other tools. In the case of Saferegex, however, it seems likely that many of the vulnerable verdicts are false positives.  

On the other hand, the tools Rescue and Revealer tend to reach similar verdicts and share these verdicts with several other tools. This may be because both tools use a detection mechanism that is at least in part based on dynamic analysis.

\subsection{Case studies}
We selected a sample of Regexes in order to glean further insights about the mechanisms' limitations. 

First, consider the regex at the heart of vulnerability  CVE-2018-20164: 
\begin{verbatim}
((?:[A-z0-9]+|[A-z-]+ ?)?(?: the )?(?:[Ss][Pp][Ii][Dd][Ee][Rr]|
[Ss]crape|[A-Za-z0-9-](?:[^C][^Uu])[Bb]ot|[Cc][Rr][Aa][Ww][Ll])
[A-z0-9])(?:(?:[ /]| v)(\d+)(?:.(\d+)(?:.(\d+))?)?)?
\end{verbatim}

This regex is used to validate user-supplied headers in HTTPS requests. It contains several evil patterns, the most evident of which are the juxtaposed quantifiers \verb![A-z0-9]+|[A-z-]+!. This is a very typical evil pattern, which we expect should be flagged by every ReDoS detection tool. Nonetheless, only three of the tools, ReDoSHunter, Revealer and Rescue, flag this regex as vulnerable. The failure of RAT and Saferegex to detect this vulnerability is puzzling since this pattern fits the definition of vulnerable regex given in both papers.  

The regex that caused CVE-2021-23341 is as follows: 
\begin{verbatim}
 %(\s*)(?:=+ +)+=+(?:(?:\r?\n|\r)\1.+)+(?:\r?\n|\r)\1(?:=+ +)
 +=+(?=(?:\r?\n|\r){2}|\s*$) 
\end{verbatim}
This regex contains a back reference (\textbackslash 1) to the  capturing group (\textbackslash s*). It was detected as vulnerable by ReDoSHunter, Rescue, and Saferegxex, but not by RAT and Revealer. Note that the former does not handle non-regular syntactic constructs, but the latter does. 

Some obviously vulnerable regexes were deemed benign by every tool. For example, the following:  
\begin{verbatim} 
Case:%  ^(.*)-iPad/(\d+)(?:\.(\d+)|)(?:\.(\d+)|)...CFNetwork
 \end{verbatim}  
contains an  anchor followed by a greedy quantifier (a pattern judged vulnerable by ReDoSHunter) and also contains several overlapping  optional quantifiers, a pattern commonly accepted as vulnerable. Nonetheless, only Rescue returned a verdict of timeout, indicating a likely vulnerability, while Saferegex returned an undetermined verdict. All other tools failed to detect the vulnerability. 

Likewise, some regexes that contain capturing groups were found to be invulnerable by all static analysis tools, but flagged by dynamic tools. 
 \begin{verbatim} 
(1470\.net crawler|...|ZyBorg)(?:[ /]v?(\d+)(?:\.(\d+)(?:\.(\d+)|)|)|)
 \end{verbatim} 
As mentioned above, only 5 regexes were deemed vulnerable by every simple tool. Following a manual inspection, we find that all of them are rather simple, and contain overlapping quantifiers. The following is a typical example:
  \begin{verbatim}
 (\w+[\.\_\-]*)*\w+@[\w]+(.)*\w+$
  \end{verbatim} 

\subsection{Multi-Factor Metric and Correction Phase}

From the above analysis, we propose the use of a multi-factor metric to balance the risks of false positives and accurate detection.

A regex is deemed vulnerable if any of the following three criteria is met. 
It is flagged as  vulnerable by both  ReDoSHunter et Revealer;
It is flagged as vulnerable by at least two of the static or grey box tools: ReScue, RAT, SafeRegex;
It is detected as vulnerable (or provokes a timeout) with a dynamic tool and is flagged as vulnerable by a static tool. 

We arrived at these criteria experimentally after examining different versions and manually examining a sample of regexes found to be vulnerable and invulnerable. 25.51\% of regexes in the aggregation of three datasets were flagged as vulnerable when this metric was used. This metric was used as the final part of this analysis, which concerns regex correction using RegexScalpel.

For the Regex\_After dataset, 36 regexes were found to be vulnerable according to this metric. RegexScalpel was able to suggest a correction for each of them. However, for  17 of them  (52.77\%), a second pass through the automated pipeline found that they were still vulnerable, using the same multi-factor metric.  

For the RegexLib dataset, 440 regexes were deemed vulnerable. Of these, RegexScalpel suggested a correction for 339 regexes, 332 of which were adequate (75.45\%). 

The Regex\_Before dataset contains 61 vulnerable regexes. Of these, RegexScalpel was able to suggest a correction for 26 regex. Only 13 of these corrected regexes were benign, according to the multi-factor metric described above. 

These results are summarized in Table \ref{tab:correct}.

\begin{table}[]
\centering
\begin{tabular}{|l|l|l|l|}
\hline
\textbf{Dataset}       & \textbf{Vulnerable}                        & \textbf{Correction}                  & \textbf{Correction}    \\ 
               &   \textbf{(multi-factor metric)}            &            \textbf{suggested}        &  \textbf{adequate} \\ \hline
Regex\_Before & 61                                 & 26 (42\textbackslash{}.6\%) & 13 (50.0\%)           \\ \hline
Regex\_After  & 36                                 & 36(100\%)                   & 19(52,7\%)            \\ \hline
RegexLib      & 440                                & 339 (77.0\%)                & 332(97.9\%)           \\ \hline
Total         & 537                                & 401(74.6\%)                 & 364(90.7\%)           \\ \hline
\end{tabular}\label{tab:correct}
\end{table}

It is important to stress that the corrected regexes suggested by Regex Scalpel are not always semantically equivalent to the original regex. Indeed, adding an arbitrary bound to a quantifier limiting it to 500 iterations is a common modification. RegexScalpel  also commonly adds anchors (\^{} and \$) to the beginning and end of regexes.  This reality, and the discouraging results reported in Table \ref{tab:correct}, strongly  suggest that developers should not rely exclusively on regex correction tools to protect their applications against the risks of ReDoS attacks. Additional defensive strategies include using regex machine engines that are immune to this type of assault and limiting the amount of time allowed for regex verification.

\section{Conclusion} \label{sec:conclu}
In this paper, we perform an empirical analysis of ReDoS vulnerabilities, drawing distinctions between them and other types of vulnerabilities. We further evaluate the effectiveness of five ReDoS detection tools and one ReDoS correction tool using over 3000 real-life regexes. 
Our study finds that ReDoS vulnerabilities are both easy to exploit and likely to be exploited. We further find that most detection and correction tools exhibit significant lacunae.

\bibliographystyle{splncs04}
\bibliography{ref}
\end{document}